\documentclass[prl,twocolumn,superscriptaddress,amsmath,amssymb,a4,floatfix]{revtex4-1}
\usepackage{graphicx}
\usepackage{bm}
\usepackage{color}

\newcommand {\Fig}[1] {Figure~\ref{#1}}
\newcommand {\fig}[1] {Figure~\ref{#1}}   

\newcommand {\ketbra}[2] {|#1 \rangle\langle #2|}

\newcommand{\be}{\begin{eqnarray}}
\newcommand{\ee}{\end{eqnarray}}
\newcommand{\bmat}{\begin{pmatrix}}
\newcommand{\emat}{\end{pmatrix}}

\newcommand{\ptone}{$^{31}$P}

\newcommand{\sitwoeight}{$^{28}$Si}
\newcommand{\sitwonine}{$^{29}$Si}

\newcommand{\beq}{\begin{equation}}
\newcommand{\eeq}{\end{equation}}
\newcommand{\beqa}{\begin{eqnarray}}
\newcommand{\eeqa}{\end{eqnarray}}

\newcommand{\ket}[1]{\left| #1 \right\rangle}

\newcommand{\ttwo}{T$_2$}

\newcommand{\tone}{$T_1$}

\newcommand{\tonee}{$T_{\rm{1e}}$}

\begin{document}

\title{Entanglement in a Solid State Spin Ensemble}

\author{Stephanie Simmons}
\author{Richard M. Brown}
\affiliation{Department of Materials, Oxford University, Oxford, OX1 3PH, United Kingdom}

\author{Helge Riemann}
\author{Nikolai V. Abrosimov}
\affiliation{Leibniz-Institut f\"ur Kristallz\"uchtung, 12489 Berlin, Germany}

\author{Peter Becker}
\affiliation{PTB Braunschweig, 38116 Braunschweig, Germany}

\author{Hans-Joachim Pohl}
\affiliation{VITCON Projectconsult GmbH, 07743 Jena, Germany}

\author{Mike L. W. Thewalt}
\affiliation{Department of Physics, Simon Fraser University, Burnaby, BC, Canada}

\author{Kohei M. Itoh}
\affiliation{School of Fundamental Science and Technology, Keio University, Yokohama, Japan}

\author{John J.  L. Morton}
\affiliation{Department of Materials, Oxford University, Oxford, OX1 3PH, United Kingdom}
\affiliation{CAESR, Clarendon Laboratory, Oxford University, Oxford, OX1 3PU, United Kingdom}

\begin{abstract}
Entanglement is the quintessential quantum phenomenon and a necessary ingredient in most emerging quantum technologies, including quantum repeaters~\cite{bdcz}, quantum information processing (QIP)~\cite{JozsaLinden} and the strongest forms of quantum cryptography~\cite{curty04}. Spin ensembles, such as those in liquid state nuclear magnetic resonance~\cite{vandersypen2001,benchmark12}, have been powerful in the development of quantum control methods, however, these demonstrations contained no entanglement and ultimately constitute classical simulations of quantum algorithms.  Here we report the on-demand generation of entanglement between an ensemble of electron and nuclear spins in isotopically engineered phosphorus-doped silicon.
We combined high field/low temperature electron spin resonance (3.4~T, 2.9~K) with hyperpolarisation of the \ptone~nuclear spin to obtain an initial state of sufficient purity to create a non-classical, inseparable state.  The state was verified using density matrix tomography based on geometric phase gates, and had a fidelity of 98\% compared with the ideal state at this field and temperature. The entanglement operation was performed simultaneously, with high fidelity, to $10^{10}$ spin pairs, and represents an essential requirement of a silicon-based quantum information processor. 
\end{abstract} 

\maketitle

Most QIP algorithms applied to spin ensembles have been implemented in a regime of weak spin polarisation. However, due to the very low purity of the states used, any exponential enhancement offered by quantum mechanics disappears when the scaling of total resources is considered. 
Highly mixed, or weakly initialised, ensembles are often interpreted as the sum of a perfectly mixed component (given by a normalised identity matrix in the density matrix representation) and a smaller pure component: $\rho_{\rm true} = (1-\epsilon)\mathbb{I}/d+\epsilon\rho_0$, where $d$ is the dimensionality of the state. 
The $\mathbb{I}$ component is invariant under unitary operations and not directly observable by magnetic resonance, which produces measurements of the population \emph{differences} across allowed electron- and nuclear-spin transitions. It is therefore straightforward to ignore the maximally mixed component: this approach is called the ``pseudo-pure approximation" \cite{pseudopure}. 

There are a number of entanglement witnesses or monotones which are able to distinguish entangled states from (classical) separable ones~\cite{horodeckireview}. A widely used test is the positive partial transpose (PPT) criterion, which is both a necessary and sufficient test of entanglement for two coupled spin-1/2 particles~\cite{Horodecki96, peres}. Applying this test to the mixed state $\rho_{\rm true}$ above, it can be easily shown that the minimum value of $\epsilon$ which permits the overall state to be entangled is 1/3~\cite{Wei}. 
  
Typical values for $\epsilon$ in liquid state NMR and ESR (at X-band, 10~GHz, 5~K) are $\epsilon\sim 10^{-5}$ and $\epsilon\sim 10^{-2}$, respectively, well below the required threshold for the PPT test. Thus, while experiments performed in this regime provide a valuable test-bed for techniques in entanglement generation and detection~\cite{mehring03,Takui2007,mehring08}, the states created are only \emph{pseudo}-entangled, and fully separable. (A notable exception was the use of chemical methods to generate highly polarised hydrogen spin pairs~\cite{anwar04}, though this is a single-shot experiment with limited scalability.) In order to overcome this limit, we require states of higher initial purity, and a method to measure the $\mathbb{I}$ component of the density matrix.

We follow a hybrid approach, using both the electron and nuclear spin associated with a phosphorous donor in silicon. Isolated donors in isotopically engineered semiconductors are of particular interest as they possess excellent decoherence characteristics (both electron and nuclear \ttwo~exceed seconds~\cite{quantummemory,alexeit2}), can be controlled with high-fidelity using microwave and radio-frequency pulses \cite{coherencesi, MortonFidelity} and are promising for integrating quantum technologies into conventional semiconductor devices~\cite{kane}. 

\begin{figure}[t] \centerline {\includegraphics[width=3.5in]{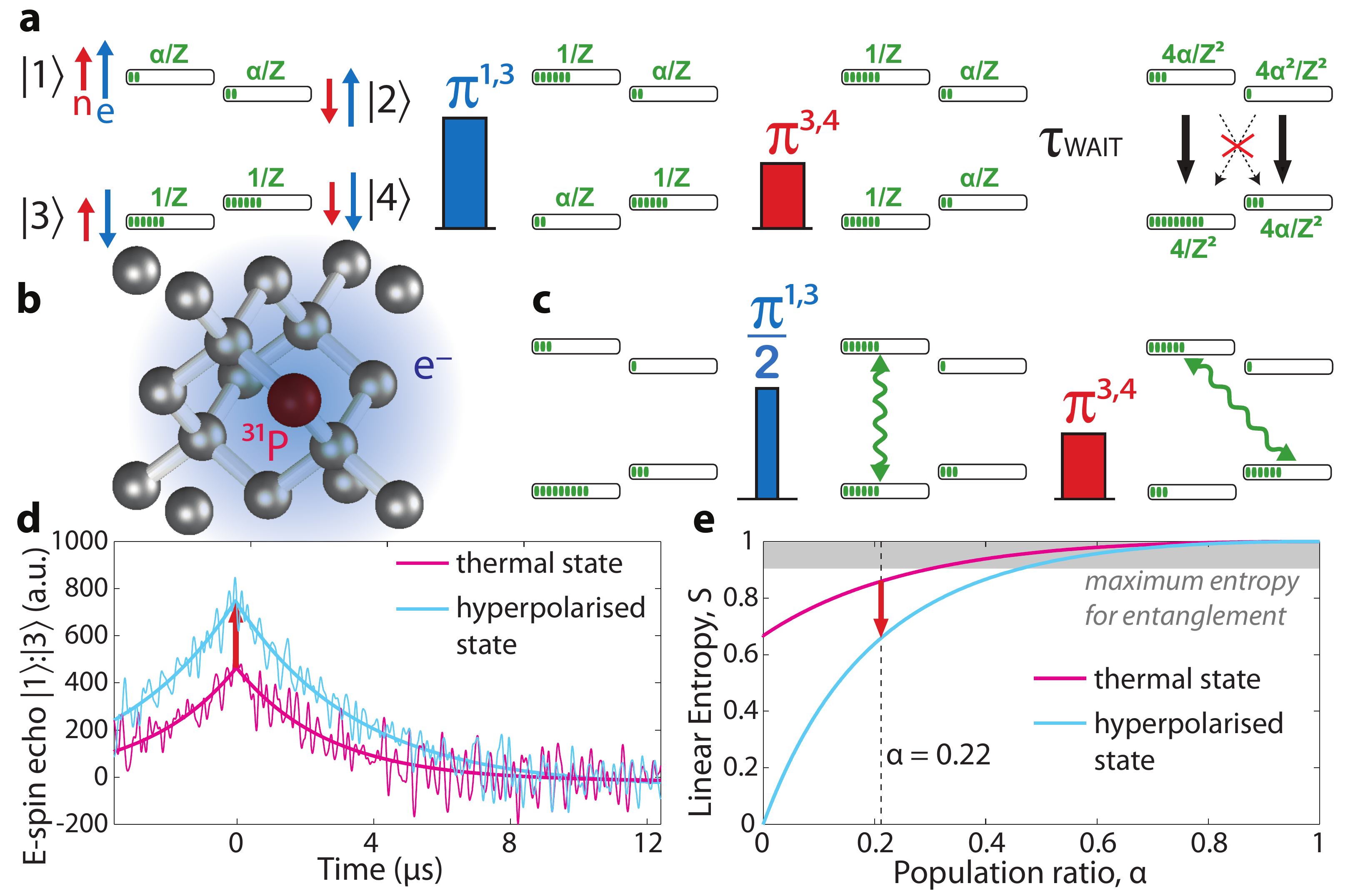}}
\caption{
Sequences for nuclear spin hyperpolarisation and entanglement generation for this coupled $S=1/2$, $I=1/2$ spin system.
a) The initial state is at thermal equilibrium where populations (shown in green) are distributed according to the electron spin polarisation at this magnetic field and temperature (see text). A pair of microwave and radiofrequency $\pi$ pulses ($\pi^{1,3}$ and $\pi^{3,4}$) are applied to move spin populations to favour the $\ket{\uparrow}$ nuclear spin state. After some time $\tau_{\rm WAIT}$, much longer than the electron spin relaxation time (\tonee), there is significant majority population in the $\ket{\uparrow\downarrow}$ state. Nuclear spin and cross-relaxation processes are much slower than \tonee. 
b) Illustration of the $^{28}$Si:P single crystal host with a substitutional phosphorus donor and its electron spin. 
c) Starting from the hyperpolarised state in (a), an electron spin coherence is generated by a microwave $(\pi/2)^{1,3}$ pulse. A radiofrequency $\pi^{3,4}$ pulse transforms this into the final entangled state, containing a superposition of  $\ket{\uparrow\uparrow}$ and  $\ket{\downarrow\downarrow}$. 
d) The growth in the electron spin echo intensity measured on the $\ket{1}:\ket{3}$ transition provides a measure of the population ratio $\alpha$. 
e) This hyperpolarisation sequence minimises the linear entropy of the two-spin state for a given value of $\alpha$.}
\label{fig:hyper} 
\end{figure}

Neglecting the weak polarisation of the nuclear spin, the initial state populations are  determined by the electron spin Zeeman energy, as shown in \fig{fig:hyper}(a) where $\alpha=\exp(-g\mu_BB/k_BT)$, $g$ is the electron g-factor, $\mu_B$ is the Bohr magneton, $k_B$ is Boltzman's constant and $B$ and $T$ are the experimental magnetic field and temperature respectively. At a high magnetic field (3.4~T) and low temperature (2.9~K), the donor electron spin is thermally polarised to $\sim 66\%$, however the \ptone~nuclear spin, with a much weaker magnetic moment, has only $\sim 0.04\%$ polarisation. Various methods exist for indirectly transferring electron spin polarisation to the nuclear spin, under the heading dynamic nuclear polarisation~\cite{solideffect,hayashi09}, and often exploiting cross-relaxation process involving simultaneous electron and nuclear spin flips. Here, we exploit the relative \emph{absence} of cross-relaxation leading to a substantial difference in the relaxation times of the electron and nuclear spin~\cite{coherencesi} in order to hyperpolarise the nuclear spin rapidly and with high efficiency. This hyperpolarisation process is similar to ``algorithmic cooling" methods, where a particular qubit relaxes quickly due to coupling to a heat bath~\cite{schulman05}.

Figure \ref{fig:hyper}(a) illustrates our method for tackling the twin challenges of measuring and minimising the $\mathbb{I}$ component in the density matrix of the coupled electron-nuclear spin system. The hyperpolarisation of the nuclear spin can be understood as a SWAP operation with the (thermally polarised) electron spin, using a combination of resonant microwave (MW) and radiofrequency (RF) $\pi$ pulses. This is followed by a delay $\tau_{\rm WAIT}$ which is substantially longer than the electron spin relaxation time \tonee~(specifically $\tau_{\rm WAIT} \approx8T_{1e}$), during which the electron spin relaxes back to thermal equilibrium. On this timescale, other relaxation processes (such as pure nuclear spin flips, or electron-nuclear spin flip-flops) are orders of magnitude slower and can be neglected. The resulting hyperpolarised state is:
\beq
\rho =\frac{4}{Z^2}(\alpha\ketbra{1}{1} + \alpha^2\ketbra{2}{2} +\ketbra{3}{3} + \alpha \ketbra{4}{4}),
\label{hyperstate}
\eeq
where $Z=2(1+\alpha)$ is a normalising constant. 

Although spin-echo sequences can only be used to probe the population differences across energy levels, we can obtain a direct measure of the population ratio $\alpha$ by measuring the electron spin-echo amplitude between levels $\ket{1}$ and $\ket{3}$ before and after the hyperpolarisation sequence, as shown in \Fig{fig:hyper}(d). Due to the enhanced polarisation of the nuclear spin, a spin-echo measured on this transition increases by a factor $2  / (\alpha + 1 )$, compared to the measurement from a fully-relaxed thermal state. This measure is strictly conservative: it places a \emph{lower} bound on the true polarisation of the electron as imperfections such as pulse errors or residual relaxation processes only lead to a lower apparent state purity. Using the measure, we observe an enhancement of the echo intensity by a factor of 1.643(2), corresponding to a upper bound of $\alpha \leq$ 0.217(2). 

This hyperpolarisation sequence corresponds to a decrease in linear spin entropy, made possible by the open quantum system's contact with the lattice heat bath (see \Fig{fig:hyper}(e)). Importantly, this approach leads to the minimum possible linear entropy given the electron spin polarisation resource and type of relaxation present~\cite{schulman05}. Entanglement is maximised in a mixed 2-qubit density matrix by first minimising the linear entropy, and then generating an entangled coherence across the levels with the greatest and \emph{second}-smallest population~\cite{Wei, Verstraete2001}. Following this strategy, we create an entangled state using a coherence-generating MW $\frac{\pi}{2}^{1,3}$ pulse, followed by an RF $\pi^{3,4}$ pulse (as shown in \fig{fig:hyper}(b)), yielding the target state:
\begin{align}
\rho = &\frac{1}{2Z^2} \bmat 1 + \alpha& 0 & 0 & 1 - \alpha\\
0& 2\alpha^2& 0 & 0\\
0& 0& 2\alpha& 0\\
1 - \alpha& 0 & 0& 1 + \alpha                 
\label{target}\emat
\end{align}
This density matrix is entangled according to the PPT criterion when $\alpha \leq 0.432 $, while other preparation methods (such as pseudo-pure state preparation) require substantially higher polarisation (see Supplementary Information).

Having prepared the initial state and performed an entangling operation, we now use density matrix tomography to extract the final two-spin state. Due to the weak magnetic moment of nuclear spins and necessarily low donor concentration in our sample, we are restricted to non-projective measurements of the electron spin ensemble along the $\sigma_x$ and $\sigma_y$ bases which can be performed selectively on the $m_I$ state of the nuclear spin (in product operator formalism, these bases can be written as $S_{x,y}I^{\alpha,\beta})$.

Diagonal elements of the density matrix (corresponding to state populations) are obtained by mapping pairs of population differences into an electron spin echo on the $\ket{1}$:$\ket{3}$ transition ($S_{x,y}I^{\alpha}$). The accurate detection of off-diagonal elements (coherences) is a more elaborate process, made by selectively labelling the coherence between each pair of eigenstates with a distinguishable time-varying phase~\cite{mehring03,HoferTPPI}. Under this process, a particular phase-accumulation rate provides the signature of a particular coherence, allowing the off-diagonal elements to be reconstructed from the amplitudes in the Fourier transform of a measured signal. 

\begin{figure}[t] \centerline {\includegraphics[width=3.5in]{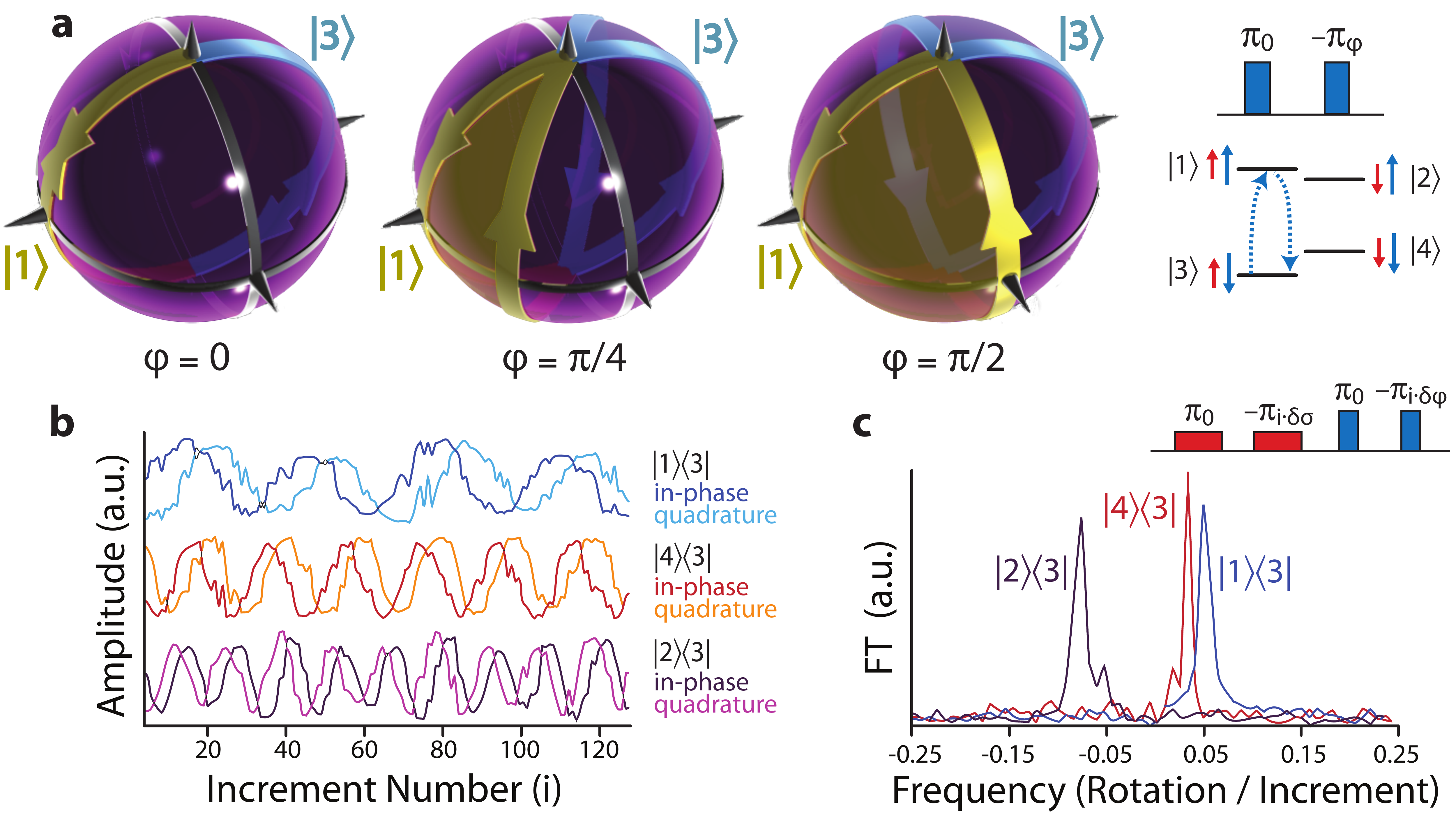}}
\caption{
Electron/nuclear spin phase rotations reveal the off-diagonal elements of the density matrix.
a)
Under the application of two consecutive $\pi^{1,3}$ pulses around different axes ($\phi$), the eigenstates $\ket{1}$ and $\ket{3}$ undergo closed trajectories on the Bloch sphere with equal and opposite solid angle $\Omega=\pm2\phi$. Each state picks up a phase equal to half this solid angle.
b) This $\pi_0-\pi_{\phi}$ phase gate is applied to both electron $\ket{1}$:$\ket{3}$ and nuclear $\ket{3}$:$\ket{4}$ transitions, where the two phases are varied by different increments $\delta\phi$ and $\delta\sigma$ as the experiment is repeated.  Example oscillations are shown for three experiments where we generate an electron coherence $\ketbra{1}{3}$, nuclear coherence $\ketbra{4}{3}$ and double quantum coherence $\ketbra{2}{3}$. 
c) Fourier transforms of the oscillations with respect to increment number show peaks located at 0.050(8), 0.031(5) and -0.079(8), in excellent agreement with the set frequencies $\nu_\phi = 2\pi/\delta\phi= 0.05$ and $\nu_\sigma = 2\pi/\delta\sigma= 0.03$. }
\label{fig:tomo} 
\end{figure}

Here we follow an approach inspired by the Aharonov-Anandan geometric phase gate~\cite{AharonovAnandan,pines88} to apply arbitrary phases in a fixed time to the four different eigenstates, and thus separately label each of the possible coherences. We apply two $\pi$ pulses, along different axes, across a transition between a pair of eigenstates. The phase acquired by each eigenstate is opposite and equal to half the solid angle of its trajectory on the Bloch sphere  (see \Fig{fig:tomo}(a)). Thus, applying $\pi^{1,3}_0$ followed by $-\pi^{1,3}_{\phi}$ (subscripts denote pulse phase and thus nominal rotation axis) leads eigenstates $\ket{1}$ and $\ket{3}$ to undergo trajectories of equal and opposite solid angle $\pm2\phi$. A similar operation is applied to the nuclear spin transition: $\pi^{3,4}_0$, $-\pi^{3,4}_{\sigma}$, such that the total operator describing the action of these four pulses is:
\beq
U(\phi,\sigma)= \bmat 
e^{-i\phi}& 0 & 0 & 0\\
0& 1 & 0 & 0\\
0& 0& e^{i(\sigma+\phi)}& 0\\
0& 0 & 0& e^{-i\sigma} 
\emat
\label{GeoZ}
\eeq

The value of $\phi$ is incremented by $\delta\phi$ on each shot of the experiment, with effective frequency $\nu_\phi = 2\pi/\delta\phi$  (and similarly for $\sigma$, $\delta\sigma$ and $\nu_{\sigma}$). 
We then map each off-diagonal element of the density matrix in turn into $S_{x,y}I^{\alpha}$ using a set of appropriate MW and RF $\pi$ pulses, and measure the amplitude of the Fourier component at the effective frequency corresponding to that coherence. Quadrature measurement allows us to discriminate between positive and negative frequencies. 
The presence of other Fourier peaks would be illustrative of pulse errors in the mapping sequence, but as seen in \Fig{fig:tomo}(c), such errors are negligible, even in the absence of operations such as phase cycling. 

\begin{figure}[t] \centerline {\includegraphics[width=3.6in]{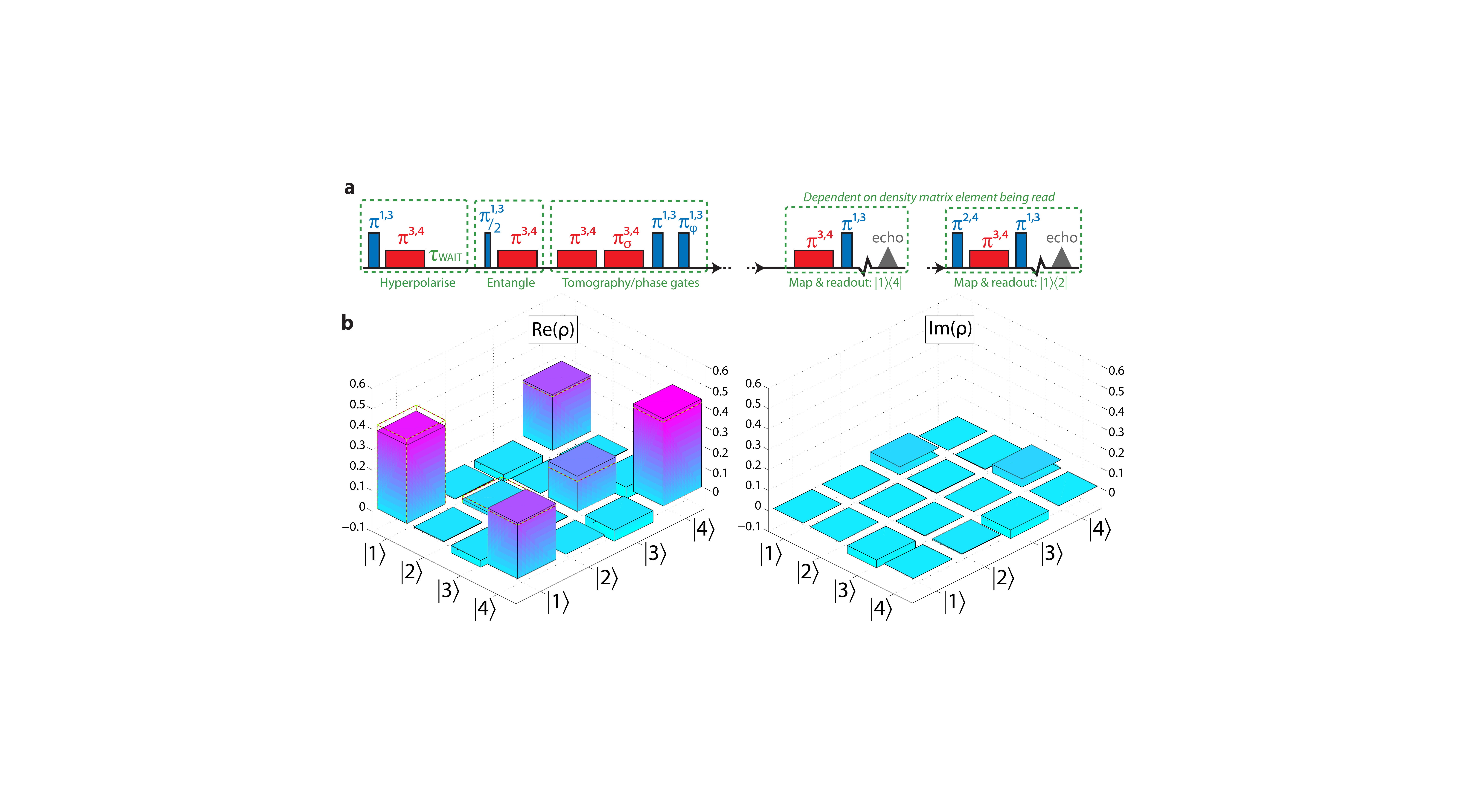}}
\caption{
Measuring an entangled density matrix. a) The full pulse sequence used to prepare, entangle and measure the two-spin state. The final readout stage was changed according to the density matrix element being measured: two examples are shown for the $\ketbra{1}{2}$ and $\ketbra{1}{4}$ states. b) The obtained density matrix is shown as solid bars, while the dashed outline (zero where not shown) shows that of an ideal state given $\alpha = 0.217$. The fidelity of the ideal state with the measured density matrix is 98\%.  }
\label{fig:results} 
\end{figure}

Combining our measurements of the identity component, diagonal, and off-diagonal elements of the density matrix of the electron-nuclear spin system, we obtain:

\scriptsize
\begin{align}
\bmat
  0.382    &         0.003 + 0.000i & -0.035 - 0.039i &  0.272      \\    
  0.003 - 0.000i&   0.017  &          -0.000 + 0.001i  & 0.001 + 0.003i\\
 -0.035 + 0.039i & -0.000 - 0.001i &  0.174    &        -0.055 - 0.042i\\
  0.272     &        0.001 - 0.003i & -0.055 + 0.042i  & 0.427 \nonumber
\emat
\end{align}
\normalsize
   
This state has a minimum eigenvalue under the PPT test of $-0.19(1)$ and a concurrence ($C$) of $0.43(4)$, each of which confirm the presence of finite entanglement. The fidelity of the measured density matrix with the target state given $\alpha = 0.217$ is $98.2(2)\%$, and results of this tomography process are shown in \fig{fig:results}.  To obtain the uncertainty in these values we used Monte Carlo generation of physical density matrices based on the standard error of each matrix element due to noise (see Supplementary Information).

The finite entanglement shown can offer direct advantages over classical methods in applications such as quantum sensors~\cite{cat9,cat13}. To achieve higher purity entangled states one could move to lower temperatures e.g.\ we would expect $C\sim0.99$ if these experiments were performed at 0.8~K. Complementary to this approach,  entanglement purification could be performed using a larger Hilbert space at each node~\cite{earlspaper}, for example using a donor atom with a higher nuclear spin (such as bismuth with $I=9/2$~\cite{georgesibi,Morley2010}).

The electron-nuclear spin entanglement generated here could also be mapped into an entangled state between nuclear spin pairs. By SWAPping the state of the electron spin with a second, coupled nucleus, for example, one could attain nuclear-spin entanglement in a regime where the thermal polarisation of the nuclei would be orders of magnitude too small and the direct coupling between them weak. Clusters of up to eight nuclei coupled to a single electron spin have been explored in other materials~\cite{spinbusmehring}, though the scaling of such an approach appears limited. A scalable network of entangled nuclear spins could be generated by exploiting the ability to ionise the donor and transfer the electron onto a neighbouring donor site~\cite{skinner03, andresen07}. These operations, combined with single-shot readout of the P-donor spin~\cite{morello10} and globally-controlled electron-nuclear spin entanglement such as we have demonstrated, form the basis for a cluster-state quantum computer based in silicon~\cite{MortonCluster}.

We thank Joe Fitzsimons, Simon Benjamin, Andrew Briggs, Alexei Tyryshkin, Steve Lyon and Brendon Lovett for valuable discussions, and P. H\"ofer and Bruker Biospin for support with instrumentation. Three-dimensional images were created using POV-Ray open-source software. 
We thank EPSRC for supporting work at Oxford through CAESR (EP/D048559/1) and the Oxford-Keio collaboration through the JST-EPSRC SIC Program (EP/H025952/1). Work at Keio has been supported by Grants-in-aid for Scientific Research by MEXT , FIRST by JSPS, Nanoquine and Keio GCOE. SS is supported by the Clarendon Fund, JJLM is supported by the Royal Society. 


\onecolumngrid
\appendix
\pagebreak

\section{Supplementary information}

\subsection{Materials and Methods}
Si:P consists of an electron spin S=1/2 (g = 1.9987) coupled to the nuclear spin $I=1/2$ of $^{31}$P through an isotropic hyperfine coupling of $a=4.19$~mT [1].
The W-band EPR signal comprises of two lines (one for each nuclear spin projection $M_I = \pm 1/2$). Our experiments were performed on the low-field line of the EPR doublet corresponding to $M_I=1/2$. At 2.9~K and 3.36~T, the electron and nuclear spin \tone~were measured to be approximately 0.6~s and 100~s, respectively. 

The sample consists of a \sitwoeight-enriched single crystal about 0.5~mm in diameter with a residual \sitwonine~concentration of order 70 ppm, produced by decomposing isotopically enriched silane in a recirculating reactor to produce poly-Si rods, followed by floating zone crystallisation. Phosphorus doping of $\sim10^{14}$ cm$^{-3}$ was achieved by adding dilute PH$_3$ gas to the Ar ambient during the final float zone single crystal growth.  Further information on the sample growth has been reported elsewhere~\cite{avagadro}.

Pulsed EPR experiments were performed using a W-band (94~GHz) Bruker Elexsys 680 spectrometer, modified to allow microwave phase control, and equipped with a 6~T superconducting magnet and a low temperature helium-flow cryostat (Oxford CF935). The cryostat was pumped to achieve a temperature of 2.88~K (internal thermocouple) with a consistent upper temperature limit confirmed by the spin temperature measurement (see text). Typical pulse times were 56~ns for a mw $\pi$ pulse and 100~$\mu$s for an rf $\pi$ pulse. To achieve arbitrary phase control, RF pulses were generated using a Rohde and Schwarz AFQ100B together with an Amplifier Research 500~W amplifier.  

\subsection{Hyperpolarisation}

Si:P is described by an isotropic spin Hamiltonian (in angular frequency units):
\begin{equation}\label{Hamiltonian}
\mathcal{H}_0=\omega_e S_z - \omega_I I_z + a \!\cdot\! \vec{S} \!\cdot\!
\vec{I},
\end{equation}
where $\omega_e=g\beta B_0/\hbar$ and $\omega_I=g_I\beta_n B_0/\hbar$ are the electron and nuclear Zeeman frequencies, $g$ and $g_I$ are the electron and nuclear g-factors, $\beta$ and $\beta_n$ are the Bohr and nuclear magnetons, $\hbar$ is Planck's constant and $B_0$ is the magnetic field applied along $z$-axis in the laboratory frame. $S_{x,y,z}=\sigma_{x,y,z}\otimes\mathbb{I}_2$ and $I_{x,y,z}=\mathbb{I}_2\otimes\sigma_{x,y,z}$. 

In the spin basis introduced in the main text, the implementation of the hyperpolarisation sequence upon the first-order Boltzmann thermal state is given by:
\begin{align}
\rho = &\frac{e^{-\hbar \omega_e  S_z / k_B T}}{\mathcal{Z}} \\
			= &\frac{1}{\mathcal{Z}}\left(\alpha\ketbra{1}{1} + \alpha\ketbra{2}{2} + \ketbra{3}{3} + \ketbra{4}{4}\right)  \nonumber\\
\pi^{1,3}_0 \rightarrow \rho = &\frac{1}{\mathcal{Z}}(\ketbra{1}{1} + \alpha\ketbra{2}{2} + \alpha\ketbra{3}{3} + \ketbra{4}{4}) \nonumber\\
\pi^{3,4}_0 \rightarrow \rho = &\frac{1}{\mathcal{Z}}(\ketbra{1}{1} + \alpha\ketbra{2}{2} + \ketbra{3}{3} + \alpha\ketbra{4}{4}) \nonumber\\
\tau_{\textrm{WAIT}} \rightarrow \rho = &\frac{4}{\mathcal{Z}^2}(\alpha\ketbra{1}{1} + \alpha^2\ketbra{2}{2} + \ketbra{3}{3} + \alpha\ketbra{4}{4}) \nonumber
\end{align}
where $\alpha = e^{-\hbar \omega_e / k_B T}$ and ${\mathcal{Z}} = 2(1+\alpha)$, $T$ is temperature in Kelvin and $\omega_e$ the resonant frequency of the electron spin. The ratio of spin-echo amplitudes between the hyperpolarised state to the thermal state on the $\ket{1}$:$\ket{3}$ transition is therefore given by $4/\mathcal{Z}=2/(\alpha + 1)$.

This hyperpolarisation process minimises the linear entropy of the system subject to the available physical processes~\cite{schulman05}.  Many measures can be used to characterise a state's mixedness; in this work we will use linear entropy because its upper limit for entangled states is tighter than the von Neumann entropy limit~\cite{Wei}. The linear entropy is given by 
\begin{align}
S_L(\rho) \equiv \frac{\mathcal{N}}{\mathcal{N}-1} [1-\textrm{trace}(\rho^2)]
\end{align}
and so the linear entropies of the thermal and hyperpolarised states are given by
\begin{align}
S_L(\rho_{\textrm{thermal}}) = \frac{2(1+4\alpha + \alpha^2)}{3(1+\alpha)^2} \nonumber\\
S_L(\rho_{\textrm{hyperpol}}) = \frac{16\alpha (1+\alpha + \alpha^2)}{3(1+\alpha)^4} \nonumber
\end{align}
The linear entropy decrease is:
\begin{align}
S_L(\rho_{\textrm{thermal}}) - S_L(\rho_{\textrm{hyperpol}}) = \frac{2(1-{\alpha})^2(1 + \alpha^{2})}{3(1+\alpha)^4}
\end{align}

\subsection{Entanglement Thresholds}
As discussed in the text, our target entangled state was optimally chosen within the limits of the state's linear entropy. We can apply the entangling sequence to the hyperpolarised state expressed in the form of a density matrix in the $\{\ket{1},\ket{2},\ket{3},\ket{4}\}$ basis. 
\begin{align}
\rho = &\frac{4}{\mathcal{Z}^2} \bmat \alpha & 0& 0& 0\\
0& \alpha^2& 0 & 0\\
0& 0& 1& 0\\
0& 0& 0& \alpha          
\nonumber\emat\\
\left(\frac{\pi}{2}\right)^{1,3}_{\pi/2},\pi^{3,4}_0 \rightarrow \rho = &\frac{2}{\mathcal{Z}^2} \bmat 1 + \alpha& 0 & 0 & 1 - \alpha\\
0& 2\alpha^2& 0 & 0\\
0& 0& 2\alpha& 0\\
1 - \alpha& 0 & 0& 1 + \alpha               
\label{target}\emat
\end{align}
The partial transpose of this matrix has a zero crossing when 
\begin{align}
 \alpha^3 - (1 - \alpha)^2/4 =0 \nonumber
\end{align}
or when $\alpha \approx 0.432$.

It is worth briefly considering the crossing-points for other preparation strategies, in each case using the optimal entangling sequence described above. Three examples include i) the pseudopure preparation scheme applied to the thermal state ii) the thermal state with no preparation and iii) the pseudopure preparation scheme applied to the hyperpolarised state. For i), the prepared pseudopure state has the form 
\begin{align}
\rho_{\textrm{pp}} = &\frac{1}{\mathcal{Z}} \bmat (1 + 2\alpha)/3& 0 & 0 & 0\\
0&(1 + 2\alpha)/3& 0 & 0\\
0& 0& 1& 0\\
0& 0 & 0& (1 + 2\alpha)/3               
\nonumber\emat
\end{align} This initial state produces the maximal amount of entanglement when arranged in the form equivalent under local unitaries to
\begin{align}
\rho_{\textrm{pp}} = &\frac{1}{\mathcal{Z}} \bmat (4 + 2\alpha)/6& 0 & 0 & (2 - 2\alpha)/6\\
0&(1 + 2\alpha)/3& 0 & 0\\
0& 0& (1 + 2\alpha)/3& 0\\
 (2 - 2\alpha)/6& 0 & 0& (4 + 2\alpha)/6               
\nonumber\emat
\end{align}
The partial transpose of this matrix has a zero crossing when 
\begin{align}
2\alpha  + \alpha^2 \nonumber =0
\end{align}
and so the only `physical' zero crossing occurs at $\alpha = 0$. Of course, these calculations assume that the nuclear spin polarisation is negligible; in the regime of $\alpha \approx 0$ the polarisation of the nucleus would have to be considered in this calculation. This would allow for a small but nonzero threshold for $\alpha$, dependent upon the nuclear spin isotope. 

For ii), the maximally entangled state is equivalent under local unitaries to
\begin{align}
\rho_{\textrm{pp}} = &\frac{1}{\mathcal{Z}} \bmat (1 + \alpha)/2& 0 & 0 & (1 - \alpha)/2\\
0&\alpha& 0 & 0\\
0& 0& 1 & 0\\
 (1 - \alpha)/2& 0 & 0& (1 + \alpha)/2               
\nonumber\emat
\end{align}
The partial transpose of this matrix has a positive zero crossing when $\alpha = -3 + 2 \sqrt{2}$, so entanglement is possible below $\alpha \approx 0.17$.

For iii), the maximally entangled state is equivalent under local unitaries to
\begin{align}\rho_{\textrm{pp}} = 
&\frac{4}{\mathcal{Z}^2} \bmat (3 + 2\alpha + \alpha^2)/6& 0 & 0 & (3 - 2\alpha - \alpha^2)/6\\
0&\alpha(2 + \alpha)/3 & 0 & 0\\
0& 0& \alpha(2 + \alpha)/3 & 0\\
 (3 - 2\alpha - \alpha^2)/6& 0 & 0& (3 + 2\alpha + \alpha^2)/6 
\nonumber\emat\end{align}
The partial transpose of this matrix has a positive zero crossing when $\alpha = \sqrt{2} -1$, and so entanglement is possible below $\alpha \approx 0.4142$.

\subsection{Density Matrix Calculations}
\begin{figure}[t] \centerline
{\includegraphics[width=2.6in]{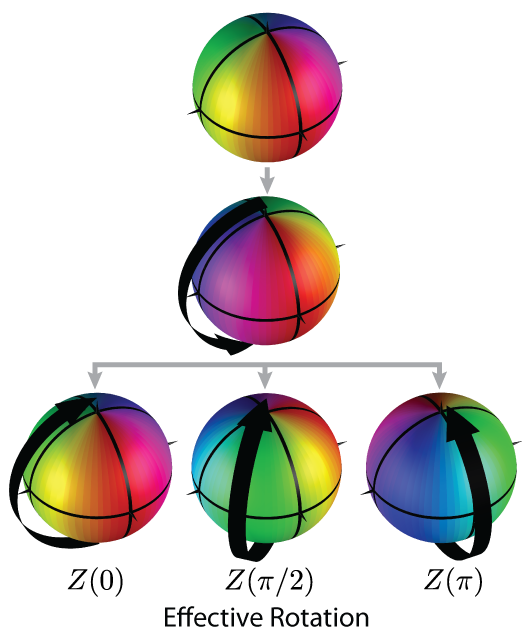}}
\caption{
\textbf{Geometric Phase Rotations} Geometric Z-gates can be applied to a single spin using two consecutive $\pi$ pulses along different axes. The black arrows at each step illustrate the applied rotation to the Bloch sphere, and the colouring displays this rotation. Three sample geometric Z-gate rotations are shown along the bottom.  }
\label{fig:z} 
\end{figure}

We used phase ($Z$) rotations to probe the existence of coherences without disturbing the state populations, which were measured separately by mapping pairs of population differences into the $S_x$ observable. In typical electron spin resonance experiments, phase rotations are not directly available operations;  they can be implemented with various strategies including the use of off-resonant pulses and appropriate delays, phase-shifted final reference frames~\cite{mehring03} and composite rotations made up of pure $X$ and $Y$ rotations of variable length~\cite{pines88}. In this work, we follow an approach inspired by the Aharonov-Anandan geometric phase gate~\cite{AharonovAnandan} and apply two selective MW or RF $\pi$ pulses of differing phases. Given the larger Hilbert space of this two-spin system, and the fact that we are applying selective rotations that cannot be understood in terms of the independent manipulation of either spin, the most instructive way to appreciate the effect of this gate is to examine the phase acquired by each eigenstate under the rotation (see Figure 2 of the main text). However, for a single spin $1/2$, the same gate has a convenient visual representation that is illustrated in \fig{fig:z} in terms of the rotation of the Bloch sphere. 

To calculate the final density matrix we first constructed the pseudo-pure density matrix. A baseline measurement was taken as an average of 2000 samples, and all datasets were baseline-corrected before processing. The population differences were measured by an average of 100 samples and scaled with respect to a measured thermal amplitude (also taken as an average over 100 samples), and adjusted to have unit trace with the addition of an appropriately scaled identity matrix.

The coherence-measurement pulse sequences were all individually tested by verifying their output frequencies when applied to their target coherences. Upper diagonal elements were gathered and their conjugate values populated the lower diagonal elements.
The coherences were collected with 128 points and baseline-corrected before being Fourier transformed and normalised with respect to the thermal amplitude. A narrow integral over the appropriate frequency position was made to measure the coherences. The frequencies were chosen such that all frequency peaks were well-resolved with a 128-point Fourier transform.

The calculated pseudopure matrix $\rho_{\textrm{pp}}$ was added to the appropriate amount of identity matrix $\mathbb{I}$ as determined by a spin-temperature measurement, where $e^{-\hbar\omega_s/k_b T} = \alpha \leq 0.217$. The explicit reconstruction is given by
\begin{align}
\rho_F=[ {\alpha}/(2(1+{\alpha}))] \mathbb{I} + [(1-{\alpha})/((1+{\alpha}))]\rho_{\textrm{pp}}\nonumber
\end{align}

The fidelity of the final density matrix $\rho_F$ with $\rho_T$, the target density matrix (calculated by Equation \ref{target} with $\alpha = 0.217$), was calculated as
\begin{align}
F(\rho_F,\rho_T) = \left( \textrm{Tr} \left( \sqrt{ \sqrt{\rho_T} \rho_F \sqrt{\rho_T} }\right) \right)^2
\end{align}
as proposed elsewhere~\cite{jozsaFidelity}.

The errors corresponding to each element were calculated according to the standard error of the measurement taken. For the populations this consisted of the standard error of the direct difference measurements; for the coherences this consisted of the standard error of the Fourier-transformed signal over the integral peak width. These density matrix element errors were transformed into final negativity, concurrence and fidelity errors by Monte Carlo generation of density matrices. The generated matrices deviated from the measured matrix in each element by an amount chosen randomly from a normal distribution whose standard deviation matched that elements' error. Once re-normalised, unphysical matrices were discarded and statistics on physical matrices were collected. In total, $2^{12}$ matrices were used to compile the final errors.

\end{document}